\documentclass[12pt]{article}
\begin{document}

\title{Hamiltonian formulation of unimodular gravity in
the teleparallel geometry}
\author{J. F. da Rocha Neto$^{1}$, J. W. Maluf$^{1}$ and S. C. Ulhoa$^{2}$\\
$^{1}$ Instituto de F\'isica, Universidade de Bras\'ilia \\
70910-900,
Brasilia, DF, Brazil.\\
$^{2}$ Instituto de Ci\^encia e Tecnologia,\\ 
Universidade Federal dos Vales do Jequitinhonha\\
e Mucuri,   39100-000, Diamantina, MG, Brazil.}
\date{\today}
\maketitle

\begin{abstract}
In the context of the teleparallel equivalent of general relativity
we establish the Hamiltonian formulation of the unimodular theory of
gravity. Here we do not carry out the usual $3+1$ decomposition
of the field quantities in terms of the lapse and shift functions,
as in the ADM formalism. The corresponding Lagrange multiplier is the
timelike component of the tetrad field. 
The dynamics is determined by the Hamiltonian constraint ${\cal
H}'_0$ and a set of primary constraints.  The constraints are first
class and satisfy an algebra that is similar to the algebra of the
Poincar\'e group. 
\end{abstract}

\noindent Keywords: Unimodular gravity, Torsion tensor, 
Hamiltonian approach \par
\noindent PACS numbers: 04.20.Fy; 04.20.Cv\par
\noindent (1) rocha@fis.unb.br\par
\noindent (2) wadih@unb.br\par
\noindent (3) sc.ulhoa@gmail.com\par

\bigskip

\section{Introduction}

The unimodular theory of gravity, or simply unimodular relativity,
is an alternative theory of gravity considered by Einstein in 1919
\cite{A1} in the cosmological context, in order to allow homogeneous,
static solutions of the fields equations. It turned out that it is
equivalent to general relativity with the cosmological constant
appearing as an integration constant. Anderson and Finkelstein in 1971
placed this theory in the Lagrangian form \cite{AF}.  The unimodular
theory of gravity is a modification of general relativity in the sense
that now it is introduced a condition that requires the determinant of
the space-time metric to have a fixed value ($g = -1$) \cite{AF}. This
condition has the effect of reducing the symmetry group from the full
space-time diffeomorphism invariance to invariance under only
diffeomorphisms that preserve the nondynamical fixed volume element.
When we introduce this condition in the Hilbert-Einstein
action, the field equations that arise are equivalent to those
obtained from Einstein's theory in the presence of a
cosmological term.

In view of the relation between the unimodular theory of gravity and
the emergence of a cosmological constant, recently this theory has been
considered from several points of view \cite{BY, Un, HT, We, Le, EA, JE}
in order to attempt a solution to the cosmological constant problems both
at the classical and quantum levels. These approaches reveal two important
aspects of the theory. First, since this theory has a fixed determinant of
the metric tensor, contributions to the energy-momentum tensor of the form
$Cg_{\mu\nu}$, where $C$ is a constant, are not sources of curvature in the
field equations \cite{Le}. This seems to solve one of the cosmological
constant problems, which is suppressing the huge contribution to the
cosmological constant that arises from quantum corrections \cite{Le}.

Second, in the ordinary (ADM type) canonical formulation of the unimodular
theory the lapse function $N$ is no longer an independent variable, since now
it is given by $N = [{g^{(3)}}]^{-1/2}$. A consequence of this
change of status of $N$ is that the primary Hamiltonian constraint of the
ordinary canonical formulation of general relativity, ${\cal H}_{\bot} = 0$,
obtained by independent variation of the total Hamiltonian with respect to
$N$, no longer emerges in the unimodular relativity as a secondary constraint,
hence the total Hamiltonian of the theory does not vanish.
The Hamiltonian constraint equations ${\cal H}_i =0$ do remain present,
because they are obtained from variation of the total Hamiltonian with respect
to the shift function $N^{i}$, which remains an independent variable. Because
of this feature of the Hamiltonian formulation of the unimodular theory of
gravity, in the procedure of canonical quantization it is possible to unfreeze
the time-dependent Schr\"odinger equation \cite{Un, HT, Le, JE}.

It is well known that for any physical theory the Hamiltonian formulation
reveals important aspects of the theory, and serve as a starting point for the
process of canonical quantization. The Hamiltonian formulation distinguishes the
hyperbolic field equations (evolution equations) from the elliptic field
equations (constraints). In the work of Arnowitt, Deser and Misner (ADM)
\cite{ADM} the Hamiltonian analysis of Einstein's general relativity reveals
that the time evolution of the field quantities is determined by the
Hamiltonian and vector constraints. Thus four of the ten Einstein's
equations acquire a well defined meaning. This is an essential feature of the
canonical quantization program.

The theory of general relativity can also be formulated in the
teleparallel (Weitzenb\"ock) geometry \cite{Wei}. In this framework
the dynamical field quantities are the tetrad fields $e_{a\mu}$,
where $a$ and $\mu$ are $SO(3,1)$ and space-time indices, respectively.
By using these fields it is possible to construct the Lagrangian
density of the teleparallel equivalent of general relativity (TEGR)
\cite{Mo, FW, Hay, JM1, JW} which generates Einstein's equations
in terms of the tetrad fields.
The Lagrangian density, in the TEGR, is given in terms of a quadratic
combination in the torsion tensor
$T_{a\mu\nu} = \partial_\mu e_{a\nu} - \partial_\nu e_{a\mu}$, which is
related to the anti-symmetric part of the Weitzenb\"ock connection
$\Gamma^{\lambda}\,_{\beta\gamma} = e^{a\lambda}\,\partial_{\beta}e_{a\gamma}$.
This connection describes the space-time endowed with absolut paralelism
\cite{JA}. The curvature tensor constructed from this connection vanishes
identically.

In the Weitzenb\"ock space-time two vectors located at $x^\mu$ and $x^\mu + dx^\mu$,
$V^\mu(x)$ and $V^\mu(x+dx)$, are said to be parallel if their
projections on the tangent space by means of the tetrad field are identical
\cite{Mo}. The vectors $V^{a}(x) = e^{a}\,_{\mu}(x)V^{\mu}(x)$ and
$V^{a}(x +dx) = e^{a}\,_{\mu}V^{\mu}(x) + (e^{a}\,_{\mu}\partial_\lambda V^{\mu} +
V^{\mu}\partial_\lambda e^{a}\,_\mu)dx^{\lambda} =
V^{a}(x) + e^{a}\,_{\mu}(\nabla_\lambda
V^{\mu})dx^\lambda$, where the covariant derivative $\nabla$ is constructed out
of the Weitzenb\"ock connection $\Gamma^{\lambda}\,_{\beta\gamma} =
e^{a\lambda}\,\partial_{\beta}e_{a\gamma}$, are projected at
$x^\mu$ and $x^\mu + dx^\mu$,
respectively. The condition of absolute paralelism, $V^a(x) = V^{a}(x+dx)$, holds
if the covariant derivative  $\nabla_\lambda V^{\mu}$ vanishes. Given that
$\nabla_\lambda e_a\,^\mu \equiv 0$, the tetrad fields $e_a\,^\mu$ constitute a set
of autoparallel fields.

In the Hamiltonian formulation of the TEGR it is possible to address
the notion of energy-momentum and angular momentum of the
gravitational field \cite{JW2}. Here, the total Hamiltonian is given by a
combination of first class constraints. The field equations of the theory, either
in Lagrangian or in Hamiltonian form, suggest definitions for the
gravitational energy-momentum and angular momentum. The Lagrangian field equations
also allow the definition of the gravitational energy-momentum tensor, as well as
the balance equations for the energy and momentum of the field. These important
aspects of TEGR serve as motivation to consider the unimodular theory of gravity in
this geometric framework. The possible relation between the cosmological constant
and dark energy constitutes a further motivation for the present investigation.
Within the context of the TEGR it will be possible to analyze whether dark energy
is an unexpected form of gravitational energy that arises as a consequence of the
cosmological constant.

In this work we present the
Hamiltonian formulation of the unimodular theory of gravity in
the context of the TEGR.  We perform the 3+1 decomposition
and obtain the total Hamiltonian
as a combination of first class constraints.  The analysis presented
here is similar to that obtained in \cite{JW1}, the difference residing in the
fact that here we introduce the unimodular condition $\sqrt{-g}- 1 = 0$
in the total Lagrangian density. As a consequence, the theory and 
in particular the Hamiltonian density depend on a cosmological constant. 
If we ultimately require the cosmological constant
to vanish, we recover the same results  presented in \cite{JW1}. In addition, we
will present the constraint algebra in a much more simple form
than that presented in \cite{JW1}. We consider this latter result as a major
achievement of the present analysis. The simplification of the
Hamiltonian formulation is crucial for a better understanding of the theory.

Notation: space-time indices $\mu, \nu, ...\;$ and SO(3,1) indices $a,
b, ...\;$ run from 0 to 3. Time and space indices are indicated
according to $\mu=0,i,\;\;a=(0),(i)$. The tetrad field is denoted by
$e^a\,_\mu$, and the flat, Minkowski space-time metric tensor raises
and lowers tetrad indices and is fixed by $\eta_{ab}=e_{a\mu}
e_{b\nu}g^{\mu\nu}= (-1,+1,+1,+1)$. The determinant of the tetrad
field is represented by $e=\det(e^a\,_\mu) = \sqrt{-g}$
and we use the constants $G=c=1$.

\section{Lagrangian Formulation}
\noindent

In this section we will first demonstrate the equivalence of the TEGR
with Einstein's general relativity. It is well known that in the
Riemannian geometry the Christoffel symbols $ ^{0}\Gamma^{\lambda}\,_{\mu\nu}$
are symmetric in the lower indices and therefore the corresponding torsion
tensor vanishes. However, in the TEGR the field equations are constructed
out of the torsion tensor $ T^{\lambda}\,_{\mu\nu}$, related to the
anti-symmetric part of the Weitzenb\"ock connection,
$\Gamma^{\lambda}\,_{\mu\nu} =
e^{a\lambda}\partial_{\mu}e_{a}\,_{\nu}$, where
$T^\lambda\,_{\mu\nu}=e_a\,^\lambda T^a\,_{\mu\nu}$ and

\begin{equation}
T^{a}\,_{\mu\nu} = \partial_\mu e^{a}\,_{\nu} -
\partial_\nu e^{a}\,_{\mu} \;,\label{1.1}
\end{equation}
The torsion-free Levi-Civita connection is given by

\begin{equation}
^0\omega_{\mu ab} = -\frac{1}{2}e^{c}\,_{\mu}(\Omega_{abc} -
\Omega_{bac} - \Omega_{cab})\;, \label{1.2}
\end{equation}
where
$$\Omega_{abc} =
e_{a\nu}(e_{b}\,^{\mu}\partial_{\mu}e_{c}\,^{\nu} - e_{c}\,^{\mu}
\partial_{\mu}e_{b}\,^{\nu})\,.$$
The Christoffel symbolos $ ^0\Gamma^{\lambda}\,_{\mu\nu}$ and the Levi-Civita
connection $^{0}\omega_{\mu ab}$ are identically related by
\begin{equation}
^0\Gamma^{\lambda}\,_{\mu\nu} = e^{a\lambda}\partial_\mu e_{a\nu} +
e^{a\lambda} (^0\omega_{\mu ab})e^{b}\,_{\nu}\;. \label{1.3}
\end{equation}
Using the above equation it is possible to obtain the identity
\begin{equation}
^0\omega_{\mu ab} = -K_{\mu ab}\;, \label{1.4}
\end{equation}
where $K_{\mu ab} = \frac{1}{2}e_{a}\,^{\lambda}e_{b}\,^{\nu}
(T_{\lambda\mu\nu} + T_{\nu\lambda\mu} + T_{\mu\lambda\nu})$ is
the contortion tensor. This identity is important in the
construction of the Lagrangian density of the TEGR.
From Eq. (\ref{1.4}) it is possible to obtain the scalar curvature
$R(^0\omega)$, from which we can build the following identity,
\begin{equation}
eR(^0\omega) = -e\Sigma^{abc}T_{abc}
+ 2\partial_{\mu}(eT^{\mu})\;, \label{1.5}
\end{equation}
where $e$ is the determinant of the tetrad field $e^{a}\,_{\mu}$
and $T^{a} = T_{b}\,^{ba}$. $\Sigma^{abc}$ is defined by \cite{JW}
\begin{equation}
\Sigma^{abc} = \frac{1}{4}\left(T^{abc} + T^{bac} - T^{cab}\right) +
\frac{1}{2}\left(\eta^{ac}T^{b} - \eta^{ab}T^{c}\right)\;,\label{1.6}
\end{equation}
In Eq. (\ref{1.5}) both sides
are invariant under Lorentz transformations.
By eliminating the divergence term in Eq. (\ref{1.5})
we can define the Lagrangian density of the TEGR as
\begin{equation}
{\cal L}(e_{a\mu}) = -ke\Sigma^{abc}T_{abc} - {\cal L}_{M}
\; ,\label{1.7}
\end{equation}
where $k = 1/(16\pi)$ and ${\cal L}_ {M}$ represent the
Lagrangian density for the matter fields. 

%%%%%%%%%%%%%%%%%%%%%%%%%%%%%%%%%%%%%%%%%%%%%%%%%%%%%%%%%%%%%%%%%%%%%%%%%%%%%%%

Since the sum of both terms on the right hand side of Eq. (\ref{1.5}) is
invariant under local Lorentz transformations, the term 
$-ke\Sigma^{abc}T_{abc}$ alone does not display the invariance, unless the
coefficients of the local Lorentz transformations fall off to zero 
sufficiently fast at spacelike infinity, so that the divergence term 
$\partial_{\mu}(eT^{\mu})$ plays no role to the local Lorentz invariance
of the action integral \cite{Cho}. In general, under an arbitrary local Lorentz
transformation the term $-ke\Sigma^{abc}T_{abc}$ is invariant up to a total
divergence. 
%%%%%%%%%%%%%%%%%%%%%%%%%%%%%%%%%%%%%%%%%%%%%%%%%%%%%%%%%%%%%%%%%%%%%%%%%%%%%%%

The variation of ${\cal L}(e_{a\mu})$
with respect to $e^{a\mu}$ yields the fields equations. They read
\begin{equation}
e_{a\lambda}e_{b\mu}\partial_{\nu}(e\Sigma^{b\lambda\nu})
- e (\Sigma^{b\nu}\,_{a}T_{b\nu\mu} - \frac{1}{4}e_{a\mu}T^{bcd}
\Sigma_{bcd}) = \frac{1}{4k}eT_{a\mu}\;, \label{1.8}
\end{equation}
where $\delta{\cal L}_{M}/\delta e^{a\mu}\equiv eT_{a\mu}$. It is possible
to show that these field equations are equivalent to Einstein's
equations. After some algebraic manipulations we verify that the left hand
side of the field equations above are identically equal to
\begin{equation}
{1\over 2} \lbrack R_{a\mu}(e) - \frac{1}{2}e_{a\mu}R(e)\rbrack \;.
\label{1.9}
\end{equation}

From now on we will consider the Lagrangian density in (\ref{1.7})
subject to the unimodular condition.  If we want
to arrive at the field equations for the unimodular theory of
gravity, we have to vary the Lagrangian density in (\ref{1.7}) subject
to the unimodular condition $e - 1 = 0$. This can be done by using the
Lagrange multipliers method. For this purpose we add to the Lagrangian
density the field $\Lambda(x)$
that yields a field equation that is precisely the unimodular condition.
Therefore the unimodular Lagrangian density is writen as
\begin{equation}
{\cal L}'(e_{a\mu}\,,\Lambda(x)) = -ke\Sigma^{abc}T_{abc} - {\cal L}_{M}
+ \Lambda (e - 1)
\; ,\label{1.10}
\end{equation}
Except for the unimodular condition, the tetrad field $e^{a}\,_{\mu}$ is
{\it a priori} unconstrained. The field equations are obtained by varying
$ {\cal L}'(e_{a\mu}\,,\Lambda(x))$ with respect to $e^{a\mu}$ and
$\Lambda(x)$, respectively. They are given by
\begin{equation}
R_{a\mu}(e) - \frac{1}{2}e_{a\mu}R(e) - \frac{1}{2k}e_{a\mu}\Lambda(x) =
\frac{1}{2k}T_{a\mu}\;, \label{1.11}
\end{equation}
\begin{equation}
e - 1 = 0\;. \label{1.12}
\end{equation}
Taking the trace of (\ref{1.11}), we obtain $\Lambda(x)$ as
\begin{equation}
\Lambda(x) = -\frac{1}{8}\left(kR(e) + \frac{1}{2}T\right)\;\label{1.13}\,,
\end{equation}
which allows us to rewrite the field equations (\ref{1.11}) as
\begin{equation}
R_{a\mu}(e) - \frac{1}{4}e_{a\mu}R(e) = \frac{1}{2k}\left(
T_{a\mu} - \frac{1}{4}e_{a\mu}T\right)\;.\label{1.14}
\end{equation}

Since the covariant derivative of (\ref{1.9}) vanishes, it is possible
to show, with the help of (\ref{1.13}), that $\Lambda(x)$ is a space-time
independent quantity,
\begin{equation}
\frac{1}{8}\partial_{\mu}\left(kR(e) + \frac{1}{2}T\right) =
\partial_{\mu}\Lambda(x) = 0\;. \label{1.15}
\end{equation}
The right hand side of Eq. (\ref{1.14}) is invariant under the
transformation
\begin{equation}
T_{a\mu} \rightarrow T_{a\mu} + e_{a\mu}C\;,\label{1.16}
\end{equation}
where $C$ is a space-time constant. These transformations may be interpreted
as corrections to the energy-momentum tensor. Therefore the tensor
$R_{a\mu}(e)$ on the left hand side of (\ref{1.14}) is not affected by the
transformations above. In addtion, under this transformation Eq. (\ref{1.13})
yields
\begin{equation}
\Lambda \rightarrow \Lambda - \frac{1}{4}C \; . \label{1.17}
\end{equation}
Thus, by combining Eqs. (\ref{1.16}) and (\ref{1.17}),
the field equations (\ref{1.11}) are unchanged under the transformations
(\ref{1.16}). A similar result was observed in \cite{Le}
in terms of metric tensor.

\section{The Legendre transform}
\noindent

In order to obtain the Hamiltonian density we rewrite the  Lagrangian
density  ${\cal L}'(e_{a\mu}\,,\Lambda(x))$  in the form
${\cal L}'= p\dot{q} - {\cal H}'_{0}$, and identify the primary constraints.
To do this, we will not carry out the $3+1$ decomposition of the field 
quantities in terms of the lapse and shift functions.
Therefore in the following both $e_{a\mu}$  and $g_{\mu\nu}$
are space-time fields. The procedure adopted here is similar to that
presented in \cite{JW1}.

From the Lagrangian density in (\ref{1.10}) we obtain the momentum canonically
conjugated to $e_{a\mu}$. It is given by
\begin{equation}
\Pi^{a\mu} = 4ke\Sigma^{a\mu0}\;. \label{2.1}
\end{equation}
Given that $\Sigma^{abc} = -\Sigma^{acb}$, we have $\Pi^{a0} \equiv 0$.
In terms of (\ref{2.1}), the Lagrangian density (\ref{1.10}) can be 
rewritten as

\begin{eqnarray}
{\cal L}'(e_{a\mu}\,,\Lambda(x)) &=& \Pi^{ai}\dot{e}_{ai} -
\Pi^{ai}\partial_{i}e_{a0} - \frac{1}{2}\Pi^{ai}T_{a0i} -\nonumber\\
&-&ke\Sigma^{aij}T_{aij} + \Lambda(e - 1)\;,\label{2.2}
\end{eqnarray}
where the dot over $e_{ai}$ represents the time derivative. Also,
we are assuming that ${\cal L}_{M}=0$.

Before we proceed, let us consider the full expression
of $\Pi^{ai}$ in terms of the torsion tensor.  From Eqs. (\ref{2.1})
and (\ref{1.6}) it can be written as
\begin{small}
\begin{eqnarray}
\Pi^{ai}&=& ke\{g^{00}(-g^{ij}T^{a}\,_{0j} - e^{aj}T^{i}\,_{0j} +
2e^{ai}T^{j}\,_{0j})+ \nonumber \\
&+& g^{0i}(g^{oj}T^{a}\,_{0j} + e^{aj}T^{0}\,_{0j})
+ e^{a0}(g^{0j}T{i}\,_{0j} +
g^{ij}T^{0}\,_{0j})- \nonumber \\
&-& 2(e^{a0}g^{0i}T^{j}\,_{0j} +  e^{ai}g^{0j}T^{0}\,_{0j})
- g^{0k}g^{ij}T^{a}\,_{kj}+ \nonumber\\
&+& e^{ak}(g^{0j}T^{i}\,_{kj} - g^{ij}T^{0}\,_{kj}) - 2(g^{k0}e^{ai}
-g^{ki}e^{a0})T^{j}\,_{ji}\}\;. \label{2.3}
\end{eqnarray}
\end{small}
Denoting $(...)$ and $[...]$ as the symmetric and
antisymmetric parts of the field quantities, respectively,
we decopose $\Pi^{ai}$ into irreducible components,
\begin{equation}
\Pi^{ai}= e^{a}\,_{k}\Pi^{(ki)} + e^{a}\,_{k}\Pi^{[ki]}
+ e^{a}\,_{0}\Pi^{0i}\;,
\label{2.4}
\end{equation}
where
\begin{eqnarray}
\Pi^{(ki)}&=& ke\{g^{00}(-g^{kj}g^{il} + g^{ik}g^{jl})
+ g^{0k}(g^{0j}g^{il} - g^{0i}g^{jl}) \nonumber \\
&+& g^{0j}(g^{0i}g^{kl} - g^{0l}g^{ik})\}(T_{l0j}
+ T_{j0l}) +  ke\Delta^{ki}\;,
\label{2.5}
\end{eqnarray}

\begin{eqnarray}
\Delta^{ki} &=& -g^{0m}(g^{kj}T^{i}\,_{mj} + g^{ij}T^{k}\,_{mj} -
2g^{ik}T^{j}\,_{mj}) - \nonumber\\
&-&(g^{km}g^{0i} + g^{im}g^{0k})T^{j}\,_{mj}\;,\nonumber
\end{eqnarray}

\begin{equation}
\Pi^{[ki]} = -ke\{g^{km}g^{ij}T^{0}\,_{mj} - (g^{km}g^{0i} -
g^{im}g^{0k})T^{j}\,_{mj}\}\;,\label{2.6}
\end{equation}

\begin{equation}
\Pi^{0i} = -2ke\{g^{ij}g^{0m}T^{0}\,_{mj}
- (g^{0i}g^{m} - g^{00}g^{im})T^{j}\,_{mj}\}\;. \label{2.7}
\end{equation}
An important point in this analysis is that only the symmetric
components $\Pi^{(ki)}$ depend on $T_{a0j}$ ,
which contains the time derivative of the tetrad field.
The other six components $\Pi^{[ki]}$ and $\Pi^{0k}$
depend solely on $T_{aij}$. Therefore we can express only six components
of the ``velocity" fields $T_{a0j}$ in terms of the six components
$\Pi^{(ki)}$. To do this we note from Eq. (\ref{2.5}) that $\Pi^{(ki)}$
depends only on the symmetric components of $T_{a0j}$. We define
\begin{equation}
\psi_{lj} = T_{l0j} + T_{j0l}\;,\label{2.8}
\end{equation}
and substitute the above definition into Eq. (\ref{2.5}). We also define
\begin{equation}
P^{ki} = \frac{1}{ke}\Pi^{(ki)} - \Delta^{ki}\;,\label{2.9}
\end{equation}
and find that $P^{ki}$ depend only on $\psi_{lj}$,
\begin{small}
\begin{eqnarray}
P^{ki}&=& -g^{00}(g^{km}g^{ij}\psi_{mj} - g^{ki}\psi) +
(g^{ok}g^{im}g^{oj} +  \nonumber \\
&+& g^{0i}g^{km}g^{0j})\psi_{mj} - (g^{ik}g^{om}g^{0j}\psi_{mj} +
g^{0k}g^{0i}\psi)\;, \label{2.10}
\end{eqnarray}
\end{small}
where $\psi = g^{ij}\psi_{ij}$.

We can now invert $\psi_{lj}$ in terms of $P^{ki}$. After a number of
manipulations we arrive at
\begin{equation}
\psi_{lj} = -\frac{1}{g^{00}}\left(P^{ki}g_{kl}g_{ij}
- \frac{1}{2}g_{lj}P\right)\;,\label{2.11}
\end{equation}
where $P = g_{ik}P^{ik}$.

By using the definition of $\Sigma^{abc}$ in terms of the torsion tensor, 
and using Eqs. (\ref{2.3}), (\ref{2.8}) and (\ref{2.11}), we conclude that
the third and fourth terms on the right hand
side of Eq. (\ref{2.2}) can be rewritten as
\begin{small}
\begin{eqnarray}
-\frac{1}{2}\Pi^{ai}T_{aoi} - ke\Sigma^{aij}T_{aij} &=&
\frac{ke}{4g^{00}}
\Big(g_{ik}g_{jl}P^{ik}P^{kl}- \frac{1}{2}P^{2}\Big) \nonumber\\
&-&  \Big(
\frac{1}{4}g^{ik}g^{jl}T^{a}\,_{ij}T_{akl} +
ke\frac{1}{2}g^{jl}T^{k}\,_{ij}T^{i}\,T_{kl} -\nonumber\\
&-&g^{il}T^{j}\,_{ij}T^{k}\,_{kl}\Big)\;. \label{2.12}
\end{eqnarray}
\end{small}
Thus, finally we obtain the primary Hamiltonian density,
${\cal H}'_{0} = \Pi^{ai}\dot{e}_{ai} - {\cal L}'$, as
\begin{equation}
{\cal H'}_{0} = {\cal H}_{0} - \Lambda(e - 1)\;,\label{2.13}
\end{equation}
where
\begin{small}
\begin{eqnarray}
{\cal H}_{0}(e_{ai}, \Pi^{ai}, e_{a0}) &=&
-e_{a0}\partial_{i}\Pi^{ai} -
\frac{ke}{4g^{00}}\Big(g_{ik}g_{jl}P^{ik}P^{kl}-\frac{1}{2}P^{2}\Big)\\ 
\nonumber
&+& ke\Big(
\frac{1}{4}g^{ik}g^{jl}T^{a}\,_{ij}T_{akl} +
\frac{1}{2}g^{jl}T^{k}\,_{ij}T^{i}\,T_{kl} -     
g^{il}T^{j}\,_{ij}T^{k}\,_{kl}\Big) \;. \label{2.14}
\end{eqnarray}
\end{small}

Now we can write the total Hamiltonian density. For this
purpose we have to indentify the primary constraints. They are
related to expressions (\ref{2.6}) and (\ref{2.7}),
which represent relations between $e_{ai}$ and the momenta
$\Pi^{ai}$. Thus we define

\begin{eqnarray}
\Gamma^{ik} &=& -\Gamma^{ki} = (\Pi^{ik} - \Pi^{ki}) +
2ke\{g^{im}g^{kj}T^{0}\,_{mj} - \nonumber \\
&{}&-(g^{im}g^{ok} - g^{km}g^{0i})T^{j}\,_{mj}\}\,, \nonumber \\
\Gamma^{0k}& =& \Pi^{0k} + 2ke\{g^{kj}g^{0m}T^{0}\,_{mj} -
(g^{0k}g^{0m} - g^{00}g^{km})T^{j}\,_{mj}\}\;.\label{2.15}
\end{eqnarray}
Before we write the total Hamiltonian density, we will simplify
the constraints above.  Since
$\Pi^{a0}\equiv 0$, we can write the constraints above
as a single constraint $\Gamma^{ab} = -\Gamma^{ba}$, where
$\Gamma^{ik} = e_{a}\,^{i}e_{b}\,^{k}\Gamma^{ab}$ and
$\Gamma^{0k} = e_{a}\,^{0}e_{b}\,^{k}\Gamma^{ab}$.
Thus in view of Eq. (\ref{2.1}) $\Gamma^{ab}$ can be written as

\begin{equation}
\Gamma^{ab} = 2\Pi^{[ab]} + 4ke(\Sigma^{a0b} -
\Sigma^{b0a})\;.
\label{2.16}
\end{equation}

Therefore the total Hamiltonian density is given by

\begin{equation}
{\cal H}' = {\cal H}'_{0} + \lambda_{ab}\Gamma^{ab}
+ \lambda_{a}\Pi^{a0}\;, \label{2.17}
\end{equation}
where $\lambda_{ab} = -\lambda_{ba}$ and $\lambda_{a}$
are Lagrange multipliers to be determined.
Although in the usual Hamiltonian formalism of the TEGR
the term that  involves the constraint $\Pi^{a0}\equiv 0$
does not generate any additional information, here we have to add it to
the  total Hamiltonian density because it will be important
to analize the time evolution of the unimodular condition.

\section{Secondary constraints}
\noindent

Considering Eq. (\ref{2.1}) we notice that the momenta
$\Pi^{a0}$ vanish identically, and so they constitute
primary constraints whose time evolution induces
secondary constraints,

\begin{equation}
C'^{a} \equiv  \frac{\delta {\cal H}'}{\delta e_{a0}}
= 0\;.\label{3.0}
\end{equation}
According to the terminology of Dirac, secondary constraints
are relations between the fields and momenta which must be
independent of the primary constraints, otherwise these
relations will be equivalent to primary constraints \cite{Dirac}.
In what follows, in order to obtain the expression of $C'^{a}$
we have to vary only ${\cal H}'_{0}$ with respect to $e_{a0}$,
because the variation of $\Gamma^{bc}$ with respect
to $e_{a0}$ vanishes identically,
\begin{equation}
\frac{\delta \Gamma^{ab}}{\delta e_{c0}} \equiv 0\;.\label{3.1}
\end{equation}
To obtain the expression of $C'^{a}$ we make use of the
variation ${\delta e^{c\mu}}/{\delta e_{a0}} =
-e^{a\mu}e^{c0}$. In addition, we need of the variation of
$P^{ij}$ with respect to $e_{a0}$. It is given by

$$\frac{\delta P^{ij}}{\delta e_{a0}} = - e^{a0}P^{ij}
+ \gamma^{aij}\;,$$ where $\gamma^{aij}$ is defined as
\begin{eqnarray}
\gamma^{aij} &=& - e^{ak}[g^{00}(g^{jm}T^{i}\,_{km} +
g^{im}T^{j}\,_{km} + 2g^{ij}T^{m}\,_{mk})+\nonumber\\
&+&g^{0m}(g^{0j}T^{i}\,_{mk} + g^{0i}T^{j}\,_{mk}) -
2g^{0i}g^{0j}T^{m}\,_{mk}+\nonumber\\
&+&(g^{jm}g^{0i} + g^{im}g^{0j} -
2g^{ij}g^{0m})T^{0}\,_{mk}]\;,\nonumber
\end{eqnarray}
which satisfies $e_{a0}\gamma^{aij} = 0$. With these considerations
we can now calculate $C'^{a}$. After a long calculation we arrive at
the expression for $C'^{a}$, which is given by
\begin{footnotesize}
\begin{eqnarray}
C'^{a}&=& -\partial_{i}\Pi^{ai} +
e^{a0}[-\frac{1}{4g^{00}}ke(g_{ik}g_{jl}P^{ij}P^{kl} -
\frac{1}{2}P^{2}) +\nonumber \\
&+& ke(\frac{1}{4}g^{im}g^{nj} T^{b}\,_{mn}T_{bij} +
\frac{1}{2}g^{nj}T^{i}\,_{mn}T^{m}\,_{ij}-
g^{ik}T^{m}\,_{mi}T^{n}\,_{nk})]- \nonumber\\
&-& \frac{1}{2g^{00}}ke(g_{ik}g_{jl}\gamma^{aij}P^{kl}
- \frac{1}{2}g_{ij}\gamma^{aij}P)-
kee^{ai}(g^{0m}g^{nj}T^{b}\,_{ij}T_{bmn}+\nonumber \\
&+&g^{nj}T^{0}\,_{mn}T^{m}\,_{ij} + g^{0j}T^{n}\,_{mj}T^{m}\,_{ni} -
2g^{0k}T^{m}\,_{mk}T^{n}\,_{ni} -
\nonumber\\
&-&2g^{ik}T^{0}\,_{ij}T^{n}\,_{nk}) - e^{a0}\Lambda e\;. \label{3.2}
\end{eqnarray}
\end{footnotesize}

The constraint above admits a simplification. After a number of
manipulations we can show that the expression above can be written as
\begin{equation}
C'^{a} =\frac{\delta {\cal H}'_{0}}{\delta e_{a0}} =
e^{a0}({\cal H}_0 - \Lambda e) + e^{ai}{\cal H}_i\;,\label{3.3}
\end{equation}
where $H_i$ is defined as
$${\cal H}_i = -e_{ci}\partial_{k}\Pi^{ck} - \Pi^{ck}T_{cki}\;.$$
From Eq. (\ref{3.3}) we note that $C'^{a}$ satisfies the following
relation
\begin{equation}
e_{a0}C'^{a} = {\cal H}_{0} -\Lambda e\;.\label{3.4}
\end{equation}
In addition, because the variation of ${\cal H}_i$ with respect to $e_{a0}$
is identically null, it follows from Eqs. (\ref{2.13}) and (\ref{3.3}) that
\begin{equation}
\frac{\delta C'^{a}}{\delta e_{c0}} 
= e^{a0}C'^{c} - e^{a0}C'^{c} \equiv 0\;,\label{3.5}
\end{equation}
Therefore, in view of Eqs. (\ref{2.13}), (\ref{2.17}) and (\ref{3.4})
we can write the total Hamiltonian density as
\begin{equation}
{\cal H'}(e_{ai}, \Pi^{ai}, e_{a0}, \lambda_{ab},\Lambda)
= e_{a0}C'^{a} + \lambda_{ab}\Gamma^{ab} + \lambda_{a}\Pi^{a0}
+ \Lambda\;, \label{3.6}
\end{equation}
in terms of the constraints $C'^{a}$, $\Gamma^{ab}$ and $\Pi^{a0}$.
We note that the vanishing of the constraints $C'^{a}, \Gamma^{ab}$ 
and $\Pi^{a0}$ does not imply the vanishing of ${\cal H}'$, which
depend on $\Lambda$.

The variation of ${\cal H'}$ with respect to $e_{a0}$
yields the constraints $C'^{a}$. Therefore we observe
that $e_{a0}$ in the total Hamiltonian density ${\cal H}'$
arises as Lagrange multipliers, together with
$\lambda_{ab}$ and $\lambda_{a}$. Moreover,
as we will see in the next section, no new constraint appears
in the formalism by time evolution of the secondary constraints
$C'^{a}$. The main difference between the formalism presented here and 
the Hamiltonian formulation presented in Ref. \cite{JW1} is that in the 
present case the canonical Hamiltonian density ${\cal H}'_{0}$ does not
vanish as a consequence of the secondary constraint $C'^{a} = 0$
(see Eq. (\ref{3.4})). This feature takes place here because of
the unimodular condition $e-1=0$, which implies that not all
components of $e_{a\mu}$ are independent. We remark that we have
not explicitly implemented in the expressions above the
condition $e-1=0$. The variation $\delta
{\cal H}'/{\delta \Lambda} = 0$ yields the unimodular condition
$e - 1 = 0$.

%%%%%%%%%%%%%%%%%%%%%%%%%%%%%%%%%%%%%%%%%%%%%%%%%%%%%%%%%%%%%%%%%%%%%%%%%%%%%%%%%
We remark that the structure of the Hamiltonian density given by Eq.
(\ref{3.6}) is very much different from the Hamiltonian formulation of tetrad
gravity constructed out of the scalar curvature density $eR(\,^0\omega)$,
in terms of the tetrad field and the spin connection as given by
Eq. (\ref{1.2}) (see, for instance, Ref. \cite{JW1991}). The essential 
difference between the Hamiltonian formulation derived from Eq. 
({\ref{1.7}) and those obtained out of invariants of the curvature tensor
(typically, the scalar curvature density) is that the Hamiltonian constraint
in the present case naturally emerges with a total divergence of the type
$-\partial_i \Pi^{ai}$ (the first term on the right hand side of of Eq. 
(\ref{3.2})), that gives rise to the total energy-momentum four-vector (see Eq.
(\ref{6.2}) below). In contrast, the Hamiltonian constraint in the ADM type 
formulation of tetrad theories of gravity does not display any nontrivial, 
total divergence (see Eq. (22) of \cite{JW1991}, which is very much similar to
the Hamiltonian constraint of the ADM formulation). It is possible to establish
total divergences, in the form of scalar or SO(3,1) vector densities, in 
theories constructed out of the torsion tensor, but not in metrical theories 
of gravity. 

We finally observe that the timelike component $e_{a0}$ of the tetrad field,
which stands as a Lagrange
multiplier in Eq. (\ref{3.6}), may be expressed in terms of the lapse and
shift functions as \cite{JW1991}

\begin{equation}
e^a\,_0=\eta^a N + N^i\,e^a\,_i\,,
\label{3.7}
\end{equation}
where $\eta^a=-Ne^{a0}$ is a timelike vector that satisfies

$$\eta_a e^a\,_i=0\,,\;\;\;\;\;\;\;\; \eta_a \eta^a=-1\,,$$
and whose direction may be fixed by means of a local Lorentz rotation.
Therefore the Lagrange multiplier $e_{a0}$ encompasses both the lapse and
shift functions, according to (\ref{3.7}). However, the lapse function
does not appear in the contraction 
$e_{a0}{C'}^a=N(\eta_a C'^a)+ N^i(e_{ai}C'^a)$. Considering the expression of 
$C'^a$ we easily find 

\begin{eqnarray}
N(\eta_a C'^a)&=& ({\cal H}_0-\Lambda e)-N^i{\cal H}_i \nonumber \\
N^i(e_{ai}C'^a)&=& N^i {\cal H}_i \,,
\label{3.8}
\end{eqnarray}
in agreement with (\ref{3.3}). In the expression above we have considered
$N=(-g^{00})^{-1/2}$ and $N^i=g^{0i}/N^2$. Thus the lapse function does not 
arise as a Lagrange multiplier in the Hamiltonian density.

%%%%%%%%%%%%%%%%%%%%%%%%%%%%%%%%%%%%%%%%%%%%%%%%%%%%%%%%%%%%%%%%%%%%%%%%%%%%%%%%%

\section{Lagrange multipliers and Poisson brackets}
\noindent

Before we obtain the Poisson brackets of the constraints of the
theory, we will determine the expressions for the Lagrange
multipliers $\lambda_{ab}$ and $\lambda_{a}$ that arise in ${\cal
H'}$. The Poisson brackets between two quantities $A$ and $B$ is
defined as
\begin{small}
$$\{A, B\} = \int d^3z\left(\frac{\delta A}{\delta e_{a\mu}(z)}
\frac{\delta B}{\delta \Pi^{a\mu}(z)} - \frac{\delta A}{\delta \Pi^{a\mu}(z)}
\frac{\delta B}{\delta e_{a\mu}(z)}\right)\;,$$
\end{small}
from what we can write down the time evolution equations. The first
set of Hamilton's equations is given by
\begin{small}
\begin{eqnarray}
\dot{e}_{a\mu}(x) &=& \{e_{a\mu}(x), \int d^3y{\cal
H'}(y)\}\nonumber\\
&=& \int d^3y\frac{\delta}{\delta \Pi^{a\mu}(x)}[{\cal H'}_{0}(y) +
\lambda_{bc}(y)\Gamma^{bc}(y) +\nonumber\\
&+&\lambda_{a}(y)\Pi^{a0}(y)]\,. \label{5.1}
\end{eqnarray}
\end{small}
In the equation above the dot over $e_{a\mu}$ represents the time
derivative. This equation can be worked out so that for
$\mu = 0$ we obtain

$$ \dot{e}_{a0} = \lambda_{a},$$
and for $\mu = j$,

\begin{equation}
T_{a0j} = -\frac{1}{2g^{00}}e_{a}\,^{k}\left(g_{lk}g_{jm}P^{lm} -
\frac{1}{2}g_{kj}P\right) + 2\lambda_{aj}\;,\label{5.1a}
\end{equation}
from what follows

$$T_{i0j} + T_{j0i} = \psi_{ij} = -\frac{1}{g^{00}}
\left(g_{il}g_{jm}P^{lm} - \frac{1}{2}g_{ij}P\right)\;,$$
and

\begin{equation}
\lambda_{ab}=\frac{1}{4}(T_{a0b} - T_{b0a} + e_{a}\,^{0}\,T_{00b}
- e_{b}\,^{0}\,T_{00a})\;.
\label{5.1b}
\end{equation}

Therefore the Lagrange multipliers acquire a well-defined
meaning in terms of the time derivatives of the field quantities and
consequently we can obtain an expression for $\Pi^{(ij)}$ in
terms of $\psi_{ij}$ by using equation (\ref{2.8}). The
dynamical evolution of the fields quantities is completed
with the second set of Hamilton's equations for $\Pi^{a\mu}$,

\begin{equation}
\dot{\Pi}^{a\mu}(x) = \{\Pi^{a\mu}(x), \int d^3y{\cal H}'(y)\} =
-\int d^3y\left(\frac{\delta {\cal H}'(y)}{\delta e_{a\mu}(x)}\right)\;.
\label{5.2}
\end{equation}

The calculations of the Poisson brackets of the constraints are very long,
tedious and intricate. Here we will just present the results. We first
calculate the Poisson brackets between ${\cal H}'_{0}(x)$ and
${\cal H}'_{0}(y)$, and then the Poisson brackets between ${\cal H}'_{0}(x)$
and $\Gamma^{bc}(y)$. They are given by, respectively,

\begin{equation}
\{{\cal H}'_{0}(x), {\cal H}'_{0}(y)\} = 0\;,\label{5.3}
\end{equation}

\begin{equation}
\{{\cal H}'_{0}(x), \Gamma^{bc}(y)\} = (e^{b}\,_{0}\,C'^{c} -
e^{c}\,_{0}\,C'^{b})\delta(x - y)\;.\label{5.4}
\end{equation}

By using the definition of $C'^{a}$ in Eq. (\ref{3.3}), and the
relation given by Eq. (\ref{3.5}), together with Eq. (\ref{5.3}),
it follows that
\begin{equation}
\{C'^{a}(x), C'^{b}(y)\} = 0\;. \label{5.5}
\end{equation}
For the calculation of the second Poisson bracket we again
use the definition of $C'^{a}$ in Eq. (\ref{3.3}) and the
fact that the variation of $\Gamma^{ab}$ and $C'^{a}$ with
respect to $e_{c0}$ is identically zero (see Eqs. (\ref{3.1})
and (\ref{3.3})). So, taking the variation of equation (\ref{5.4})
with respect to $e_{a0}$ on both sides we obtain
\begin{equation}
\{C'^{a}(x), \Gamma^{bc}(y)\} = \left(\eta^{ab}C'^{c}
- \eta^{ac}C'^{b}\right)\delta (x - y)\;. \label{5.6}
\end{equation}
And finally, by means of explicit calculations we obtain the third
Poisson bracket, which is given by
%\begin{footnotesize}
\begin{equation}
\{\Gamma^{ab}(x), \Gamma^{cd}(y)\} = \left(
\eta^{ac}\Gamma^{bd} + \eta^{bd}\Gamma^{ac}
- \eta^{ac}\Gamma^{bd} -
\eta^{bd}\Gamma^{ac}\right)\delta (x-y)\;.\label{5.7}
\end{equation}
%\end{footnotesize}
We remark here that the Poisson brackets of the constraints
$\Pi^{a0}$ with $C'^{a}$ and $\Gamma^{ab}$ vanish identically.

Let us now to analize the time evolution of the unimodular condition,
which amounts to calculating the time evolution of the
determinant $e$, namely,

$$\dot{e}(x) = \{e(x),\int {\cal H}'(y)d^{3}y\}\,.$$
Working out both sides of this equation and using that
$\dot{e}_{a0} = \lambda_{a}$ and $e^{aj}\lambda_{aj} = 0$ we obtain
the following relation,
\begin{footnotesize}
$$\dot{e} = ee^{aj}\dot{e}_{aj} = ee^{aj}\left[\partial_{j}e_{a0}
-\frac{1}{2g^{00}}e_{a}\,^{k}\left(g_{lk}g_{jm}P^{lm} -
\frac{1}{2}g_{kj}P\right)\right]\;.$$
\end{footnotesize}
Assuming that $e$ is not null and that $e^{aj}$ are arbitrary field
quantities, this relation is equivalent to the relation shown in Eq.
(\ref{5.1a}), that is obtained from the first set of Hamilton's
equations. Thus we see that the  unimodular condition does not
generate any additional constraint in the formalism.

Therefore, in view of the constraint algebra above for
$C'^{a}$ and $\Gamma^{ab}$, we see that these constraints
constitute a set of first class constraints. The algebra is very much
similar to the algebra of the Poincar\'e group.
As asserted at the end of the previous section,
given that the total Hamiltonian density is a combination
of the constraints $C'^{a}$, $\Gamma^{ab}$ and $\Pi^{a0}$, plus
the cosmological constant $\Lambda$, no new constraint arises in the
formalism by means of the time evolution of $C'^{a}$ and $\Gamma^{ab}$,
as the Poisson brackets (\ref{5.5}), (\ref{5.6}) and (\ref{5.7})
vanish weakly. 
It is important to note that if we make $\Lambda = 0$ in this
theory, the Hamiltonian formalism presented here reduces
to the Hamiltonian formalism of the TEGR presented in
Ref. \cite{JW1} and as a consequence all 
Poisson brackets presented in Ref. \cite{JW1} can be
obtained from the Poisson brackets shown in
Eqs. (\ref{5.5} - \ref{5.7}).

%%%%%%%%%%%%%%%%%%%%%%%%%%%%%%%%%%%%%%%%%%%%%%%%%%%%%%%%%%%%%%%%%%%%%%%%%%%%%%%%

\section{Summary of the results of the paper}
The results of the paper can be summarized as follows.\par
\bigskip
\noindent {\bf 1.} The configuration space is described by the tetrad field
$e^a\,_\mu$ and the function $\Lambda(x)$ which, in view of Eq. (\ref{1.15}),
turns out to be a constant. The Lagrangian field equations
for $e^a\,_\mu$ are given by (\ref{1.11}) or by Eq. (\ref{6.3}) below.
Note that Eq. (\ref{1.15}) is obtained by taking 
the covariant derivative of the field equations. \par
\bigskip
\noindent {\bf 2.} The total Hamiltonian density is given by Eq. (\ref{3.6}).
The phase space of the theory is constructed out of the
pairs of canonically conjugated field quantities ($e_{ai}, \Pi^{ai}$) and 
($e_{a0}, \Pi^{a0}$), and $\Lambda$. We found that it was not necessary to 
introduced the momentum $\Pi_{\Lambda}$ canonically conjugated to $\Lambda$, 
since we would have $\Pi_{\Lambda}=0$. In view of the fact that  
Lagrangian density does not contain the time derivative of $e_{a0}$, we have 
$\Pi^{a0}=0$. The complete set of Lagrange multipliers is ($e_{a0}$,
$\lambda_{ab}$, $\lambda_a$). \par
\bigskip
\noindent {\bf 3.} All constraints are first class. They are given by
$C'^a$ (Eq. (\ref{3.3})), $\Gamma^{ab}$ (Eq. (\ref{2.16})), and by the trivial
constraint $\Pi^{a0}=0$. 
In view of Eqs. (\ref{2.13}) and (\ref{2.17}) the condition $e-1=0$
follows from the equation $\delta {\cal H}'/{\delta \Lambda} = 0$.\par
\bigskip
\noindent {\bf 4.} The constraint algebra is given by Eqs. (\ref{5.5}), 
(\ref{5.6}) and (\ref{5.7}). The Poisson bracket of the quantity $(e-1)$ with
the total Hamiltonian yields ultimately the evolution equation for the tetrad
field $e_{a\mu}$, and therefore it does not generate additional constraints.
Moreover, the Poisson brackets between $\Pi^{a0}$ and $C'^a$ and $\Gamma^{ab}$
vanish strongly in view of Eqs. (\ref{3.1}) and (\ref{3.5}).\par
\bigskip
\noindent {\bf 5.} The physical degrees of freedom of the theory may be counted
in the following way. The pair of dynamical field quantities 
$(e_{ai},\Pi^{ai})$ displays $12+12=24$ degrees of freedom. The $4+6$ first 
class constraints $(C'^a, \Gamma^{ab})$ generate symmetries of the action, and 
thus they reduce 10+10=20 degrees of freedom. Therefore in the phase space of 
the theory there are 4 degrees of freedom, as expected. The unimodular 
condition $e-1=0$ enforces the diffeomorphisms of the theory to be 
{\it transverse} diffeomorphisms $x'^\mu=x^\mu +\xi^\mu(x)$, defined by the 
condition $\partial_\mu \xi^\mu=0$ \cite{EA}. Thus the unimodular 
condition $e-1=0$ reduces one degree of freedom of the tetrad field, and at 
the same time reduces the symmetry under diffeomorphisms. Therefore it 
does not alter the counting of physical degrees of freedom.

The action of the constraints $C'^a$ and $\Gamma^{ab}$ on the tetrad field
may be computed by means of the Poisson brackets defined in section 5. We find 
it more convenient to analyse separately the action of ${\cal H}'_0$ and 
${\cal H}_i$, instead of $C'^a$. Let 
$\varepsilon_{ab}(x)=-\varepsilon_{ba}(x)$, 
$\varepsilon^i(x)$ and $\varepsilon^0(x)$ represent arbitrary infinitesimal 
functions. After some calculations we find

\begin{equation}
\delta e_{a\mu}(x)\equiv
\varepsilon_{bc}(x)\int d^3 y\lbrace e_{a\mu}(x),\Gamma^{bc}(y)\rbrace=
2\varepsilon_{ab}(x) e^b\,_\mu\,,
\label{7.1}
\end{equation}

\begin{equation}
\delta e_{a\mu}(x)\equiv
\varepsilon^i(x)\int d^3 y\lbrace e_{a\mu}(x),{\cal H}_i(y)\rbrace=
-\varepsilon^i(x) \delta^k_\mu \partial_ie_{ak}\,,
\label{7.2}
\end{equation}

\begin{eqnarray}
\delta e_{a\mu}&\equiv&
\varepsilon^0(x)\int d^3 y\lbrace e_{a\mu}(x),{\cal H}'_0(y)\rbrace\nonumber\\
&=&\varepsilon^0(x)\delta^k_\mu \lbrack\dot{e}_{ak}(x)-
2\lambda_{ak}(x)\rbrack \,,
\label{7.3}
\end{eqnarray}
where $\lambda_{ak}$ is defined by (\ref{5.1b}),

$$\lambda_{ak}={1\over 4}(T_{a0k}-T_{k0a}+e_a\,^0 T_{00k})\,, $$
and 

$$ \dot{e}_{ak}=\int d^3y \lbrace e_{ak}, {\cal H}'(y) \rbrace\,.$$
Equations (\ref{7.1}) and (\ref{7.2}) indicate that $\Gamma^{bc}$ and
${\cal H}_i$ have a clear interpretation as generators of local Lorentz 
transformations and spatial diffeomorphisms, respectively. Equation
(\ref{7.3}) tells us that ${\cal H}'_0$ generates the time evolution 
of $e_{ak}$ provided the constraint $\Gamma^{bc}$ vanishes strongly, so 
that the Hamiltonian density does not contain the Lagrange multipliers 
$\lambda_{ab}=-\lambda_{ba}$.
However, in the general case, when $\Gamma^{bc}$ is not required to vanish,
we have

\begin{equation}
\delta g_{ij}=\delta(e^a\,_i e_{aj})=\varepsilon^0(x)
\lbrace g_{ij}(x), \int d^3 y {\cal H}'_0(y)\rbrace 
=\varepsilon^0(x)\dot{g}_{ij}\,,
\label{7.4}
\end{equation}
and

\begin{equation}
\delta g_{\mu\nu}=\varepsilon_{ab}(x)
\lbrace g_{\mu\nu}(x),\int d^3 y \Gamma^{ab}(y)\rbrace =0\,.
\label{7.5}
\end{equation}

%%%%%%%%%%%%%%%%%%%%%%%%%%%%%%%%%%%%%%%%%%%%%%%%%%%%%%%%%%%%%%%%%%%%%%%%%%%%%%%%

\section{Concluding remarks}

In this paper we have obtained the Hamiltonian formulation and constraint
algebra of the unimodular theory of gravity in the framework of the TEGR.
The constraints are first class, and the constraint algebra is presented in
a more simple form, as compared to the formulation obtained in Ref.
\cite{JW1}. The simplification achieved is significant. Although in the
unimodular theory of gravity the condition $e=1$ holds, we have kept the
determinant $e$ in all expressions. The transition to the ordinary
formulation of the TEGR is easily obtained by just making $\Lambda=0$ and
dropping the condition $e=1$.

In the constraint algebra determined by Eqs. (\ref{5.5}), (\ref{5.6}) and 
(\ref{5.7}) the structure functions are space-time independent functions. This
feature may be relevant to a possible approach to the quantization of the
gravitational field. We recall that the master constraint programme for loop 
quantum gravity \cite{Thiemann} is an approach to the canonical quantization of
the gravitational field whose idea is to replace the infinity of constraints of
the theory (one at each space-time event) by a single master equation. The 
difficulty in applying the master constraint 
programme to the known formulations of canonical gravity is that the
representation theory of the usual Dirac algebra of constraints 
(the hypersurface deformation algebra) is very intricate due to the
space-time dependent structure functions that arise in the Poisson brackets of
the constraints. On the other hand, the strucure of the algebra given by Eqs. 
(\ref{5.5}), (\ref{5.6}) and (\ref{5.7}) is very simple. The representation
of this algebra may lead to a viable approch to the quantization of gravity.

The field equations for the gravitational field in the Hamiltonian or
Lagrangian form of the TEGR allow definitions of the energy-momentum and
angular momentum of the gravitational field. These definitions are not obtained
out of the action integral or the total Hamiltonian. In the framework of 
unimodular relativity we establish these definitions in similarity with the 
previous approach \cite{JW2}. We consider first Eq. (\ref{3.2}). The equation
$C'^{a}=0$ may be written in a simplified form as

\begin{equation}
-\partial_i\Pi^{ai}=h^a+ e^{a0}\,\Lambda\,e\,,
\label{6.1}
\end{equation}
where the intricate definition for $h^a$ may be obtained directly from
Eq. (\ref{3.2}). In similarity with the procedure of Ref. \cite{JW2}, the
integral form of the equation above yields the definition of the total energy,
which includes now the contribution of the cosmological constant,

\begin{equation}
P^a=- \int_V d^3x\, \partial_i\Pi^{ai}\,.
\label{6.2}
\end{equation}
$V$ is a finite volume of the three-dimensional space. This definition
may also be obtained in the Lagrangian framework. The field equation
(\ref{1.11}) may be written as
\begin{small}
\begin{equation}
e_{a\lambda}e_{b\mu}\partial_{\nu}(e\Sigma^{b\lambda\nu})
- e (\Sigma^{b\nu}\,_{a}T_{b\nu\mu} - \frac{1}{4}e_{a\mu}T^{bcd}
\Sigma_{bcd}) = \frac{1}{4k}e(T_{a\mu}+e_{a\mu}\Lambda)\;. \label{6.3}
\end{equation}
\end{small}
By following the same procedure of Ref. \cite{JW5}, we find that the equation
above may be expressed in terms of $\Pi^{ai}$ according to

\begin{equation}
\partial_i \Pi^{ai}=-k\,e\, e^{a\mu}(4\Sigma^{bj0}T_{bj\mu}-
\delta^0_\mu \, \Sigma^{bcd}T_{bcd})-e\,e^a\,_\mu(T^{0\mu}
+g^{0\mu}\Lambda)\,.\label{6.4}
\end{equation}
The integral form of this equation yields

\begin{eqnarray}
P^a&=&\int_V d^3x\,e e^a\,_\mu(t^{0\mu}+T^{0\mu}+g^{0\mu}\Lambda)\nonumber \\
&=&-\int_V d^3x \partial_i \Pi^{ai}\,.
\label{6.5}
\end{eqnarray}
The quantity
$t^{\lambda \mu}=k(4\Sigma^{bc\lambda}T_{bc}\,^\mu-
g^{\lambda\mu}\Sigma^{bcd}T_{bcd})$ is a tensor under general coordinate
transformations, and is interpreted as the gravitational energy-momentum
tensor \cite{JW5,JW6}. In the absence of the energy-momentum $T^{\mu\nu}$ for
the matter fields, $P^a$ does represent the gravitational energy-momentum
four-vector, again including the contribution of the cosmological constant.
We emphasize that the definition of the energy-momentum four-vector $P^a$
is obtained directly from the field equations, not from the action integral.
%%%%%%%%%%%%%%%%%%%%%%%%%%%%%%%%%%%%%%%%%%%%%%%%%%%%%%%%%%%%%%%%%%%%%%%%%%%%%%%%%
The tetrad field $e^a\,_\mu$ yields the space-time metric tensor, and at the
same time establishes the frame for a given observer in space-time endowed
with the four-velocity $u^\mu=e_{(0)}\,^\mu$.

\end{document}